# High-throughput investigation of tunable superconductivity in FeSe films

Zhongpei Feng[1,2]*, Jie Yuan[1,2,3]*, Jun Li[4]*, Xianxin Wu[5], Wei Hu[1,2], Bing Shen[1,2], Mingyang Qin[1,2], Lin Zhao[1,2], Beiyi Zhu[1,2], Valentin Stanev[6,7], Miao Liu[1,8], Guangming Zhang[9], Huaixin Yang[1,2], Jianqi Li[1,2], Xiaoli Dong[1,2,3], Fang Zhou[1,2], Xingjiang Zhou[1,2], Fedor V. Kusmartsev[10,11,12], Jiangping Hu[1,2], Ichiro Takeuchi[6,7]†, Zhongxian Zhao[1,2,3], and Kui Jin[1,2,3]‡

[1]Beijing National Laboratory for Condensed Matter Physics, Institute of Physics, Chinese Academy of Sciences, Beijing 100190, China.

[2]University of Chinese Academy of Sciences, Beijing 100049, China.

[3]Key Laboratory for Vacuum Physics, University of Chinese Academy of Sciences, Beijing 100049, China.

[4]School of Physical Science and Technology, ShanghaiTech University, Shanghai 200031, China.

[5]Institut für Theoretische Physik und Astrophysik, Julius-Maximilians-Universität Würzburg, 97074 Würzburg, Germany.

[6]Department of Materials Science and Engineering, University of Maryland, College Park, MD 20742, USA.

[7]Center for Nanophysics and Advanced Materials, University of Maryland, College Park, MD 20742, USA.

[8]Energy Technologies Area, Lawrence Berkeley National Laboratory, Berkeley, California 94720, USA.

[9]State Key Laboratory of Low Dimensional Quantum Physics and Department of Physics, Tsinghua University, Beijing 100084, China.

[10]Department of Physics, Loughborough University, Loughborough LE11 3TU, UK.

[11]Micro/Nano Fabrication Laboratory Microsystem and THz Research Center, Chengdu, Sichuan, China

[12]ITMO University, St. Petersburg 197101, Russia

## Abstract

There is an ongoing debate about the relative importance of structural change versus doping charge carriers on the mechanism of superconductivity in Fe-based materials. Elucidating this issue is a major challenge since it would require a large number of samples where structure properties or the carrier density is systematically varied. FeSe, with its structural simplicity, is an ideal platform for addressing this question. It has been demonstrated that the superconductivity in this material can be controlled through crystal lattice tuning, as well as electronic structure manipulation. Here, we apply a high-throughput methodology to FeSe to systematically delineate the interdependence of its structural and electronic properties. Using a dual-beam pulsed laser deposition, we have generated FeSe films with a marked gradient in the superconducting transition temperature (below 2 K < $T_c$ < 12 K) across 1 cm width of the films. The $T_c$ gradient films display ~ 1% continuous stretch and compression in the out-of-plane and in-plane lattice constants respectively, triggering the continuous enhancement of superconductivity. Combining transport and angular-resolved photoemission measurements on uniform FeSe films with tunable $T_c$ from 3 K to 14 K, we find that the electron carrier density is intimately correlated with $T_c$, i.e., it increases by a factor of 6 and ultimately surpasses the almost constant hole concentration. Our transmission electron microscope and band structure calculations reveal that rather than by shifting the chemical potential, the enhanced superconductivity is linked to the selective adjustment of the $d_{xy}$ band dispersion across the Fermi level by means of reduced local lattice distortions. Therefore, such novel mechanism provides a key to understand discrete superconducting phases in FeSe.

Subject Areas: Condensed Matter Physics, Quantum Physics, Superconductivity

In Fe-based superconductors [1-3], the bulk tetragonal iron selenide (β-FeSe, space group *P4/nmm*) undergoes a structural transition to orthorhombic phase at $T_s$ ~ 90 K and becomes superconducting at $T_c$ ~ 8 K [4,5]. $T_c$ can be significantly boosted by modifying the lattice or manipulating the conduction carrier concentration. Under high pressure, $T_c$ can be pushed up to ~ 37 K, with concomitant reduction in both the in-plane (*a*) and out-of-plane (*c*) lattice parameters and a decrease in the *c/a* ratio [6,7]. By gating FeSe thin flakes, an onset transition at ~ 48 K has been observed, attributed to the dominance of electron carriers in the high-$T_c$ phase [8]. Both the structure and $T_c$ of FeSe films can be further tuned by reducing the thickness from thirty-five to one monolayer (1 ML) [9]. Compared to the bulk, the in-plane lattice parameter of 1 ML FeSe expands ~ 3% and the superconducting energy gap closes at ~ 65 K [10-12]. An angle-resolved photoemission spectroscopy (ARPES) study of the electronic structure of a FeSe single crystal has revealed a hole-like Fermi surface at the **Γ**(0, 0) point and a rather complex evolution with temperature of the band dispersion at the **M** point (along the Fe-Fe direction) in momentum space [13]. As the thickness of the film is reduced, the hole-like band at the **Γ** point is gradually pulled down from the Fermi level ($E_F$) and eventually vanishes, while the electron-like character of the Fermi surface at the **M** point becomes more pronounced [9].

The gating and ARPES experiments [14,15] have suggested that the increase of the electron carriers density due to shifting of the chemical potential is beneficial to $T_c$. In addition, pressure, which can also lead to $T_c$ increase, not only causes compression of the crystal lattice but also likely enhances the hole contribution [16]. In general, the roles of the lattice parameters variation and conduction carriers doping in $T_c$ enhancement are difficult to disentangle. In order to improve our understanding of FeSe, it is essential to systematically investigate the correlations among crystal structure, carrier density and superconductivity in this material. To solve this problem, we first systematically studied a series of high-quality *uniform* FeSe films (~ 160 nm), whose $T_c$ can be tuned from 0 K to 14 K, where the highest $T_c$ is almost double that of a pristine FeSe crystal. These samples were made by conventional pulsed laser deposition (PLD) technique on $CaF_2$ and $SrTiO_3$ substrates (Supplemental Figs. S1-S3). Importantly, they are sufficiently stable in air for ex-situ electrical transport measurements [17] and can be also cleaved for the ARPES experiment [18].

Our transport data in combination with the electronic structure extracted by ARPES uncover an unambiguous relation between the density of conduction carriers and $T_c$. That is, the increase of $T_c$ is associated with a gradual increase in the electron carrier density ($n_e$), while the hole carrier density ($n_h$) remains approximately constant. Such relation is inconsistent with the commonly accepted picture of $T_c$ enhancement, by the shifting of chemical potential, for which the hole carrier number should be greatly reduced. To further elucidate the mechanism of $T_c$ enhancement, we turned to the link between crystal lattice and band structure. However, the lattice parameter data collected from more than 1000 individual uniform films show significant scatter against $T_c$ (Supplemental Fig. S4), commonly seen in previous work [17,19]. Usually, samples with the sharpest superconducting transition (Supplemental Fig. S5) provide more accurate lattice constants, but obtaining the volume of data necessary for a reliable analysis requires significant amount of work. Alternatively, we developed a novel high-throughput PLD deposition method to grow the samples, in which natural gradient in $T_c$ and continuous change in the lattice constants give us a panoramic view of the relationship between the lattice constants and $T_c$ with high precision, especially the in-plane lattice versus $T_c$.

*Carrier property of uniform FeSe films.* We start with the transport data of uniform film samples (Fig. 1(a)), where the notation is such that, SC03 indicates a film with $T_c$ ~ 3K. The selected samples (with $T_c$'s ranging from 3 to 14 K) were patterned into Hall bars for simultaneous measurements of in-plane resistivity $\rho_{xx}$ and Hall resistivity $\rho_{xy}$ (obtained by sweeping the magnetic field at fixed temperatures). The widths of the superconducting transition in all films are less than 1 K (e.g. $\Delta T_c = 0.3$ K for the SC12, Supplemental Fig. S5). As can be seen, $\rho_{xy}$ is roughly proportional to the magnetic field ($B$) from 200 to 20 K (Figs. 1(d), 1(f) and Supplemental Fig. S6), indicating that the Hall coefficient $R_H$ ($=\rho_{xy}/B$) is field independent. The basic features of $R_H$ with decreasing temperature are: *i*) it is nonmonotonic, *ii*) it changes its sign from negative to positive at a cross-over temperature $T^* \sim 120 \pm 10$ K for all the samples, and *iii*) it gets suppressed as $T_c$ increases (Fig. 1(b)). Surprisingly, the $R_H(T)$ curves of all the samples almost overlap with each other for $T > T^*$. Note that the band splitting also occurs at the same - slightly higher than the $T_s$ – temperature [13,20]. Thus, we conclude that all the samples here exhibit electronic structure that is qualitatively similar above $T^*$, but differs considerably below $T^*$ (as attested by the diverging $R_H$ behavior). The magnetoresistivity (MR $=\rho_{xx}(B)-\rho_{xx}(0T)$) is proportional to $B^2$ and always positive, and it increases in magnitude as the temperature is reduced (Figs. 1(c), 1(e) and Supplemental Fig. S6). Such behaviors of Hall and MR are reminiscent of the two-type-carrier transport commonly observed in Fe-based superconductors [21-24].

Figure 2 illustrates the evolution of the electronic structure with $T_c$, obtained by combining the results of transport and ARPES measurements. We first extract $n_e$ and $n_h$; a commonly accepted method for doing this is to fit the isothermal MR and Hall resistivity curves simultaneously with the two-carrier semi-classical Drude-Sommerfeld model [25-27]. Such model works well for all the samples, as demonstrated by the perfect match of the experimental data (symbols) and the fits (solid lines) in Figs. 1(c)-1(f). For both electrons and holes the density decreases and the mobility increases with decreasing temperature (Figs. 2(a)-2(c), Supplemental Fig. S7). Such trend is consistent with previous report [28]. On the other hand, as $T_c$ is tuned the hole carrier density stays almost at a constant value for fixed temperature, e.g., $n_h \sim 2 \times 10^{20}$ cm$^{-3}$ at 20 K. Not coincidentally, no obvious changes are observed at $\Gamma$ near $E_F$ (Figs. 2(d)-2(f)), where the hole-like bands are resolved. On the other side, the change at the **M** point near $E_F$ is apparent. This suggests the variation of electron carrier density. Indeed, the electron carrier density changes by $\sim 6$ fold, from $n_e \sim 5 \times 10^{19}$ cm$^{-3}$ for SC03 to $\sim 3 \times 10^{20}$ cm$^{-3}$ for SC12 (Figs. 2(a)-2(c)). The major findings here on our uniform FeSe films therefore are: *i*) the same $T^* \sim 120 \pm 10$ K marks the beginning of divergence on the $R_H(T)$ curves; this temperature is close to the band splitting temperature observed in ARPES, namely the electronic nematic transition temperature [13]; *ii)* as the $T_c$ goes up, $n_e$ increases by a factor of 6 whereas $n_h$ remains roughly constant at low temperatures. For higher $T_c$ cases, $n_e$ becomes comparable to $n_h$ and even surpasses it for $T_c > 10$ K.

We emphasize that all the above measured samples were deposited from a same FeSe target, indicating that the nominal stoichiometry is fixed for all samples. Both the energy dispersive X-ray and the inductively coupled plasma techniques do not see clear variation of the compositions in their resolution limit, which suggests that the changes in the ratio of Fe to Se must be less than 1%. We assume a small amount of excess Fe in the composition, which can lead to electron doping in the compound, with ~ 0.5% increase of Fe leading to ~ 0.05 Å out-of-plane expansion of the lattice (Supplemental Fig. S8 and Note). However, the simple electron-filling picture cannot explain why the Hall coefficients for all different samples overlap with each other above $T^*$, as well as the fact that the hole carrier concentration remains almost constant at low temperatures. Subsequently, two key questions arise: what causes the significant variation of $T_c$, and does the crystal lattice change with $T_c$?

*Explicit lattice-$T_c$ library on high-throughput FeSe films.* In order to obtain a more detailed picture of the subtle variation of $T_c$ vs the lattice constants, we have looked to the occurrence of natural spread during a new high-throughput PLD growth, where local variation in the excimer laser fluence within a spot on the target can lead to a subtle distribution of composition and the crystal structure in the deposited film across a single substrate. For instance, it has been reported that varying the fluence could result in an expansion of *c*-axis accompanied by a slightly reduced ratio of Sr to (Sr+Ti) in the deposition of $SrTiO_3$ films [29,30]. The fact that the laser fluence has an important effect on the $T_c$ of FeSe films can be seen by comparing films deposited with different fluences (Supplemental Fig. S9). We have thus generated a laser-fluence spread by using two sliding laser beams to create a gentle distribution of the beam density across a focused laser spot on a single FeSe target, in order to fabricate a transferred FeSe film with modulated properties across the substrate (Figs. 3(a) and 3(b)). By carefully adjusting the density distribution of the mixed beams, we have succeeded in fabricating a natural FeSe spread sample of single (00*l*)-orientation with high crystallinity (Supplemental Fig. S10). By scanning the crystal structure across the length of the substrate (30 mm) with an X-ray micro-beam (where the beam size is ~ 0.4 mm), we were able to obtain the evolution of lattice parameters from two full sets of Bragg peaks, i.e. out-of-plane FeSe (002) and in-plane FeSe (220), which we refer to $CaF_2$ (002) and $CaF_2$ (400), respectively. Walking from position 0 to 30 mm on the substrate, the FeSe (002) peak first moves to a lower angle and then shifts back (Fig. 3(c)); in contrast, the FeSe (220) peak moves in the opposite direction (Fig. 3(d)). Simultaneously, $T_c$ was mapped from a series of resistance curves obtained by patterning the sample into an array of micro-bridges (each 0.5 mm in width along the 30 mm length direction across the insulating $CaF_2$ substrate (Fig. 3(d)).

As a result, an explicit lattice-$T_c$ library was created for the FeSe material for the first time (see Fig. 4(a)). Along the *y* direction defined in Fig. 3, the $T_c$ first increases, reaches a maximum ~ 12 K at the center and then drops to ~ 0 K at the other end. The distribution follows the symmetry of the beam density, so we choose to only describe the features in half of the sample, within coordinates from 0 to 15 mm. Following the *y* direction, the *c*-axis parameter gradually expands from 5.51 to 5.57 Å, whereas the *a*-axis parameter gradually shrinks from 3.78 to 3.73 Å. Due to a fixed Bragg diffraction peak of the single substrate as the reference substance, a resolution of ~ 0.01 Å is achieved. Consequently, the correlation between the lattice parameter and $T_c$ becomes unambiguous. This is another advantage of our high-throughput film synthesis besides the high efficiency. The clear variation of the lattice on the same substrate also helps to exclude the strain effect often considered in monolayer films. Hence, the lattice-$T_c$-carrier relationship is accomplished as follows: *i)* the electron carrier density is positively correlated with the $T_c$, while the hole carrier density remains almost constant (Fig. 4(b)); *ii)* $T_c$ can be tuned from 0 to 14 K, accompanied by an out-of-plane lattice expansion and an in-plane lattice compression (Fig. 4(c)).

*Microstructure analysis and discussion.* We further investigated the microstructure of the superconducting films by transmission electron microscopy (TEM). Figs. 5(a) and 5(b) show two typical electron diffraction patterns taken along the [100] zone axis for samples with $T_c$ ~ 3 K (a) and 9 K (b), respectively. The main diffraction spots with relatively strong intensity in the diffraction pattern are the basic Bragg reflections from the well identified tetragonal unit of FeSe. Another set of diffused reflection spots with slightly smaller *d*-space, indicating the presence of impurity domains, can be clearly seen in the low-$T_c$ sample (Fig. 5(c)); the corresponding high angle annular dark field (HAADF) image shows that the domain size is less than 5 nm on average. For the high-$T_c$ sample, both the Bragg reflection (Fig. 5(b)) and HAADF image (Fig. 5(d)) demonstrate a much purer FeSe phase. Intuitively, local lattice distortions by the impurity domains bring stress over the whole sample and modify the crystal lattice; reducing the number of these domains enhances the superconductivity.

The remaining question is the connections among the lattice modification, the evolution of the carrier concentration, and $T_c$. Including random distribution of impurity domains in band structure calculations is rather challenging; however, in principle one can still capture some key physics using the explicit lattice parameter library. The hole pockets around $\Gamma$ are known to be dominated by the $d_{xz}/d_{yz}$ orbital bands (the $d_{xy}$ band is below the Fermi level) [31], while the electron pockets around **M** are attributed to all three $t_{2g}$ orbitals [13,32]. Fig. 4(d) shows the band structure of FeSe obtained from three sets of lattice parameters by the first-principle calculations. The key findings are as follows: with decreasing $a$-axis parameter and increasing $c$-axis parameter, the most noticeable change in the electronic structures takes place in the $d_{xy}$ bands: the $d_{xy}$ bands shift up in energy around $\Gamma$ and shift down around **M**, while $d_{xz}/d_{yz}$ bands exhibit little change (Fig. 4(d)).

As the diagonal hopping between $d_{xy}$ orbitals is sensitive to Se height but $d_{xz}/d_{yz}$ is not, a small variation in Se height will only have effect on the $d_{xy}$ bands. With slight decrease in $a$ axis, the Se height increases and this leads to reduction of the coupling between Fe $d_{xy}$ and Se $p_x/p_y$ orbitals. Because the effective hopping between $d_{xy}$ orbitals from the above coupling is negative, this reduction should increase diagonal hopping for $d_{xy}$ orbitals, which, in turn, causes the $d_{xy}$ band's upshift around $\Gamma$ and downshift around **M**. Below $T^*$, the change of $d_{xy}$ band contribution does not affect conduction hole bands around the $\Gamma$ point, as the $d_{xy}$ band is below $E_F$, but increases conduction electrons around the **M** point by modulating the $d_{xy}$ band dispersion. Note that all the calculations are based on the room temperature lattice data, no dramatic change in electron carrier density is expected. Nevertheless, we conjecture that at low temperatures the adjustment of the $d_{xy}$ band dispersion is amplified, and thus more electron carriers are generated. Very recently, local orthorhombic lattice distortions persisting above the structure transition temperature have been observed in FeAs-based superconductors, linked to the proliferation of nematic fluctuations [33]. It is tempting to bridge a link between the local lattice distortions by tiny impurity domains reported here and the nematicity. For the low-$T_c$ sample, another set of diffused reflection spots indicates the distribution of impurity domains is regular, correlated with the nematicity. With increasing $T_c$, such domains are diminishing and distribute randomly, subsequently, the nematicity is suppressed. On the other hand, the increase of the $d_{xy}$ orbital on electron Fermi surfaces from the downshift of $d_{xy}$ bands was found to decrease the nematicity [34]. As the superconductivity competes with nematicity [35], the suppression of nematicity by diminishing the impurity domains can enhance the superconductivity. In order to further explore this issue, more advanced probes are needed to provide high quality temperature dependent information on FeSe films.

Overall, by controlling $T_c$ of the uniform FeSe films from 0 K to 14 K, we were able to observe a 6-fold increase of the electron carriers density above the superconducting transition, while the hole carriers density remained almost unchanged. Such intriguing observation cannot be attributed to an electron filling effect, because the shifting of chemical potential should have a strong impact on the hole carriers as well. Furthermore, an explicit lattice structure library, rapidly created on our single-chip high-throughput FeSe film with $T_c$ gradient from 2 to 12 K, revealed that the enhancement of $T_c$ is accompanied by ~ 1% continuous stretch and compression in the out-of-plane and in-plane lattice constants, respectively. Micro-structure data further uncover that local lattice distortions by tiny impurity domains cause the observed lattice evolution. These domains are diminished in sample with high $T_c$. Therefore, a self-consistent picture able to explain the enhanced superconductivity emerges: the lattice modulation increases the role of the $d_{xy}$ bands around the **M** point, which contribute more conduction electrons in higher-$T_c$ sample. Hence, the unusual carrier density features in FeSe films are due to the selective adjustment of the $d_{xy}$ band dispersion across the Fermi level. This mechanism unrevealed in previous work is distinct from the scenario of $T_c$ enhancement by shifting the chemical potential, which therefore may provide a key to understand discrete superconducting phases in FeSe [8,36]. Our work also demonstrates the effectiveness of combining a high-throughput strategy [37-39] with conventional individual experiments in order to uncover subtle, yet important clues to understanding the mechanism of superconductivity.

We would like to thank K. Liu, Y.F. Yang, J.G. Cheng, R.M. Fernandes, and T. Xiang for fruitful discussions, and L.H. Yang for technical support. This work was supported by the National Key Basic Research Program of China (2015CB921000, 2016YFA0300300, 2016YFA0300301, 2017YFA0302902, 2017YFA0302904, 2017YFA0303003, and 2018YFB0704100), the National Natural Science Foundation of China (11674374, 11474338, 11574372, 11334010, and 61771234), the Key Research Program of Frontier Sciences, CAS (QYZDY-SSW-SLH001 and QYZDY-SSW-SLH008), the Strategic Priority Research Program of CAS (XDPB01, XDB07020300, XDB07020100 and XDB07030200), the Beijing

[*]These authors contributed equally to this work.

[†]Corresponding author.

takeuchi@umd.edu

[‡]Corresponding author.

kuijin@iphy.ac.cn

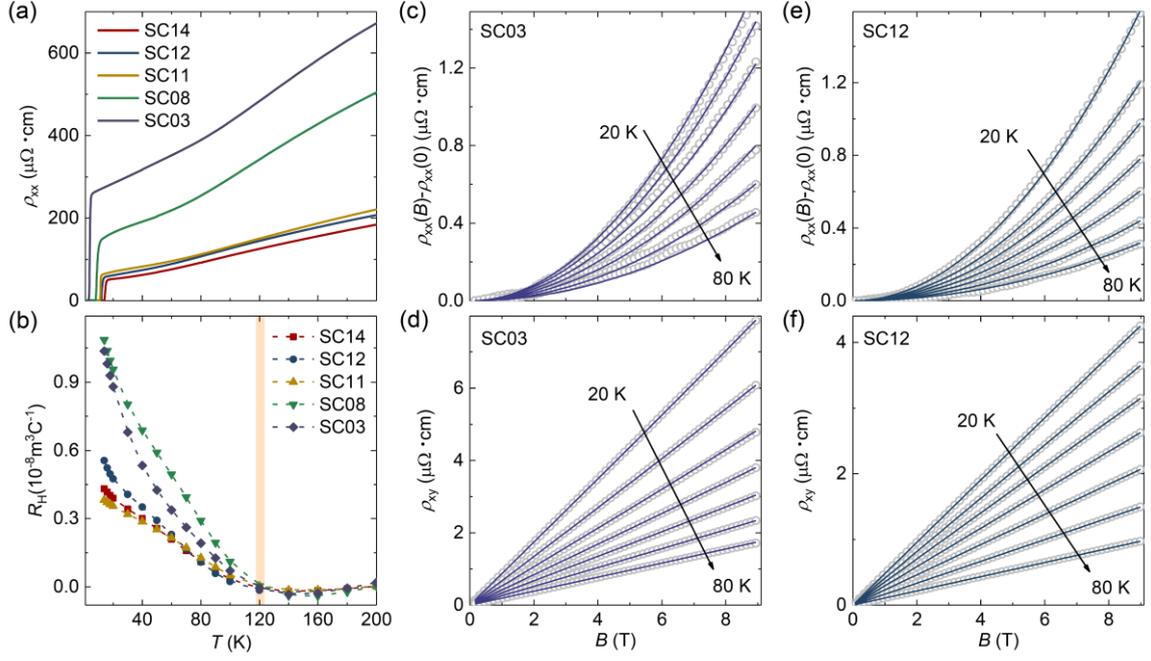

FIG. 1. Electrical transport properties of uniform FeSe films. (a) The temperature dependence of in-plane resistivity for uniform FeSe films with different $T_c$'s (3 K, 8 K, 11 K, 12 K, and 14 K, denoted as SC03, SC09, SC11, SC12, and SC14, respectively). (b) The corresponding temperature dependent Hall coefficients [$R_H(T)$] for the above samples. All the $R_H(T)$ curves across the zero at around $T^* \sim 120 \pm 10$ K (the vertical shadow area). (c)-(f) The isothermal magnetoresistivity [MR = $\rho_{xx}(B) - \rho_{xx}(0)$] and Hall resistivity [$\rho_{xy}(B)$] at 10 K intervals for two samples SC03 and SC12. Here, the symbols are experimental data and the solid lines are the fits simultaneously to MR and Hall resistivity with the two-carrier semi-classical Drude-Sommerfeld model as described in Supplemental Note 3.

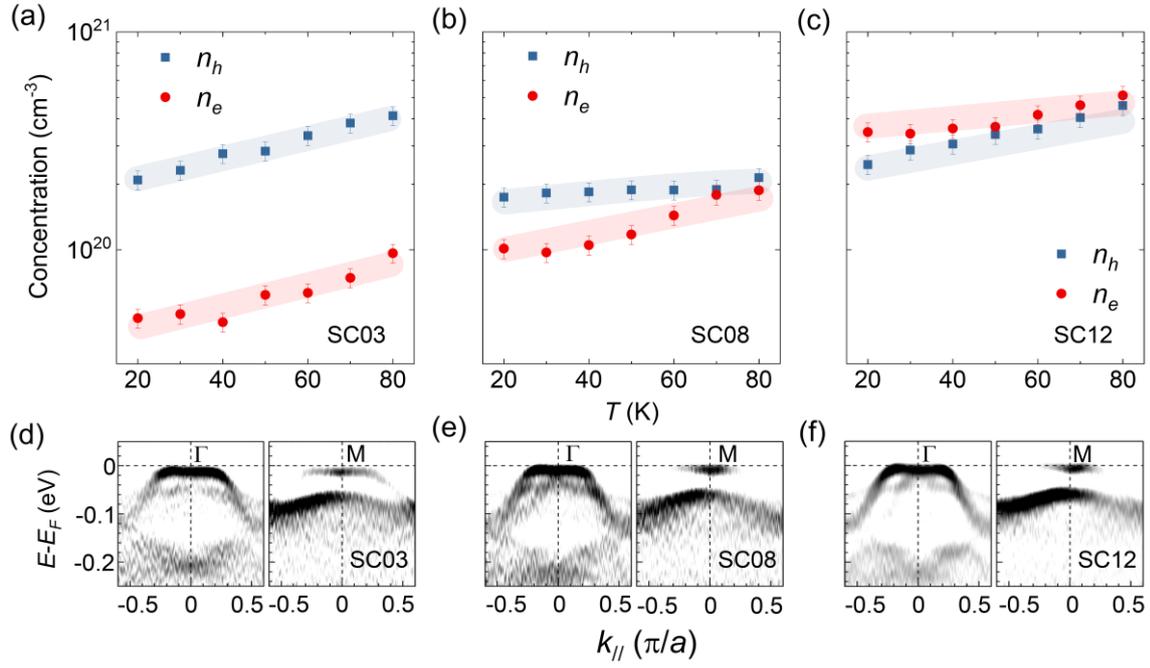

FIG. 2. Evolution of conduction carries and electronic structure. (a)-(c) The electron and hole concentrations as a function of temperature for SC03, SC08, and SC12. The carrier densities are estimated from the Hall measurements (Supplemental Note 3 and Fig. S8). (d)-(f) Band structure of the three samples at 30 K obtained from the ARPES measurements. Left and right panels in each figure correspond to the energy dispersion cuts across **Γ** and **M**, respectively.

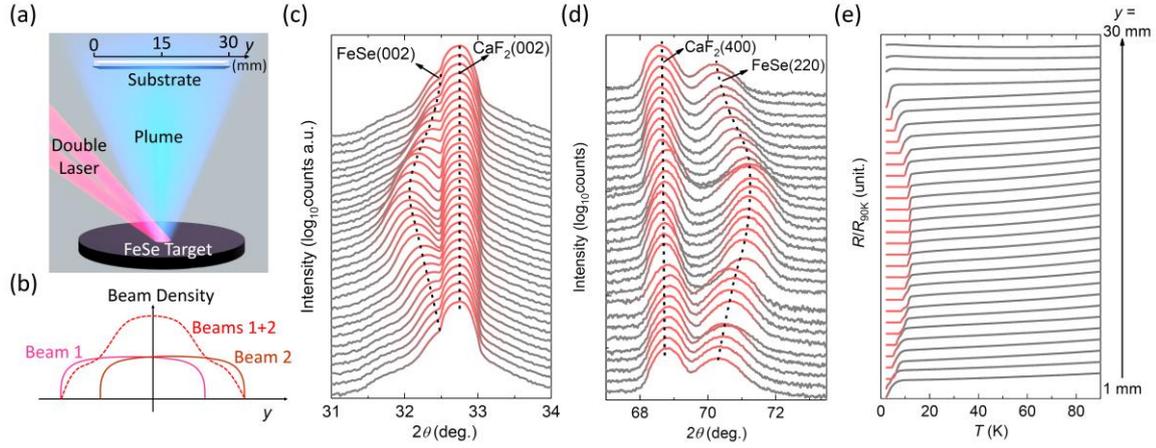

FIG. 3. High-throughput FeSe films and fast characterizations. (a) Schematic process of the double-beam pulsed laser deposition focused onto the target with controllable displacement. (b) A diagram illustrating the single beam density in a rectangle shape and double-beam density in a trapezium-like pattern along the *y* direction. Combination of two shifted laser beams can create a gentle distribution of laser fluence (dotted line). (c) The X-ray diffraction patterns for the out-of-plane (002) peak of the high-throughput film along the *y* direction from 0 to 30 mm. A clear shift referred to the (002) peak of the single $CaF_2$ substrate is observed, first to lower angle approaching the middle (*y* ~ 15 mm) and then back. (d) The in-plane FeSe (220) peak along the same direction, which moves in an opposite way. (e) The temperature dependence of normalized resistance along the *y* direction, with the highest $T_c$ in the middle corresponding to the largest *c* axis as well as the shortest *a* axis.

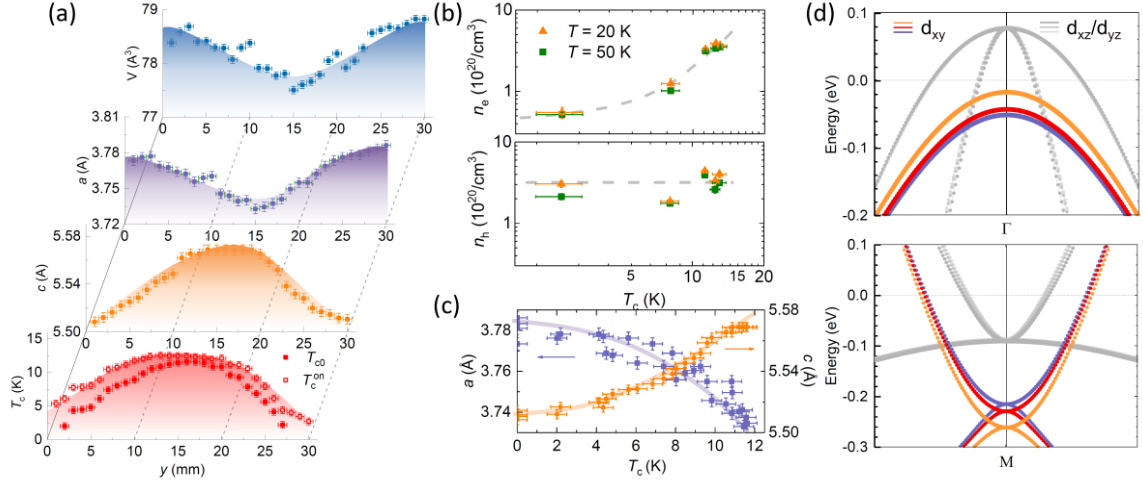

FIG. 4. Correlations among the $T_c$, carrier densities and lattice parameters. (a) The location (*y*-direction) dependence of $T_c$, *a* axis, *c* axis and unit cell volume (*V*) of high-throughput FeSe films. (b) Upper panel: $n_e$ versus $T_c$ of different samples; Lower panel: $n_h$ versus $T_c$. The plots are in log-log scale, and the lines are to guide the eye. (c) The lattice parameters *a* and *c* as a function of $T_c$. The superconductivity is gradually enhanced accompanied with continuous stretch in the *c* axis and compression in the *a* axis. (d) Band structures around **Γ** and **M** points calculated from our experimental crystal lattice data. Here we show band structures based on three sets of lattice constants (Supplemental Note 4): $a_{purple} > a_{red} > a_{orange}$ and $c_{purple} < c_{red} < c_{orange}$. The size of circles denotes the weight of the Fe $d_{xy}$ orbitals. The $d_{xy}$ orbitals are below the $E_F$ around **Γ**, but pass through the $E_F$ around **M**. Hence, the dichotomy between the electron and hole conduction carriers results from an orbital-selective effect. In conjunction with the transport data below $T^*$, the lattice modulation will bring considerable influence on the $d_{xy}$ bands rather than the $d_{xz}/d_{yz}$ bands at the Fermi surface.

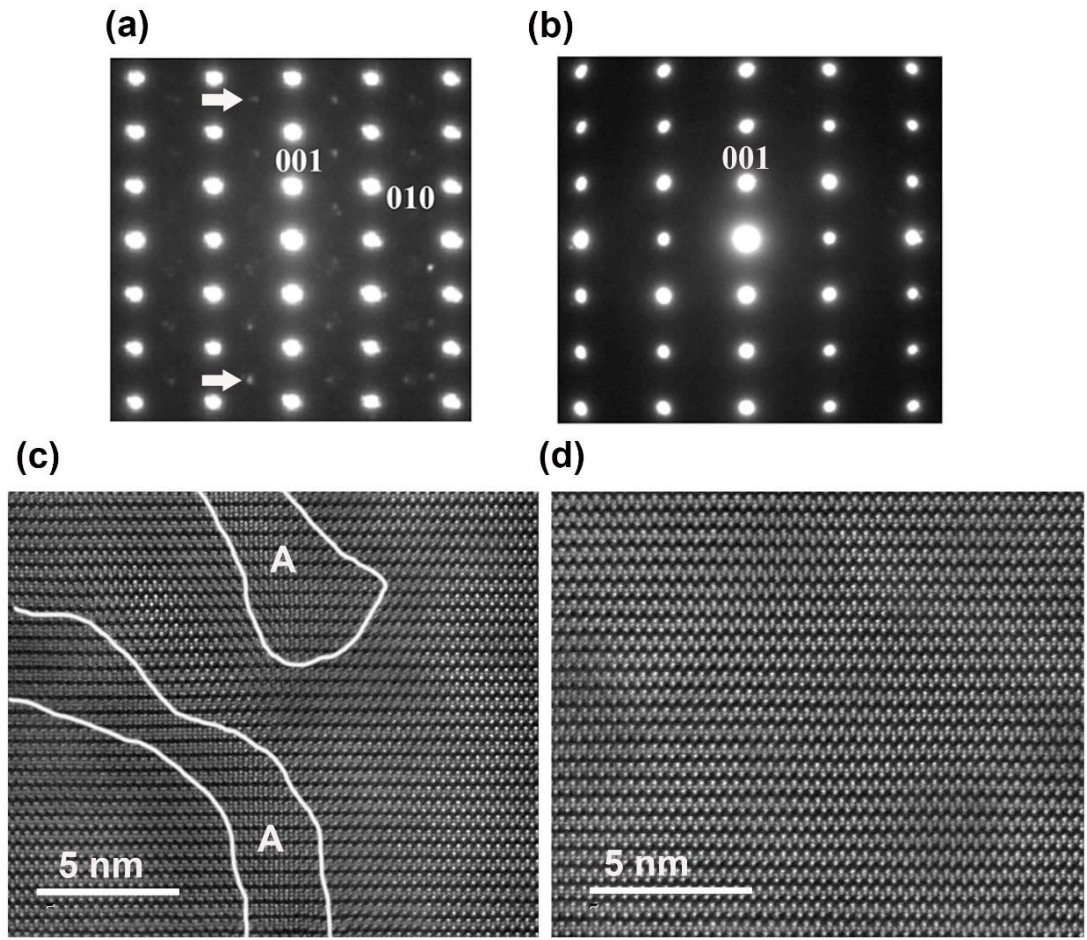

FIG. 5. Electron diffraction patterns and the corresponding HAADF STEM images, taken along the [100] zone axis for two superconducting FeSe films. (a) and (c) TEM data for thin film with $T_c$ of ~ 3 K. Additional weak Bragg spots are indicated by arrow and the circled areas (**A**), indicate the appearance of the impurity domains in the sample. (b) and (d) Corresponding TEM data for thin film with $T_c$ of ~ 9 K.